\documentclass[lettersize,journal]{IEEEtran}
\usepackage{amsmath,amsfonts}
\usepackage{algorithm}
\usepackage{algorithmic}

\usepackage{array}
\usepackage[caption=false,font=normalsize,labelfont=sf,textfont=sf]{subfig}
\usepackage{textcomp}
\usepackage{stfloats}
\usepackage{url}
\usepackage{verbatim}
\usepackage{lipsum}
\usepackage{graphicx}
\usepackage{epstopdf}

\usepackage{cite}

\usepackage{geometry}



\begin{document}

\title{Fair AI-STA for Legacy Wi-Fi: Enhancing Sensing and Power Management with Deep Q-Learning}

\author{Peini Yi,~\IEEEmembership{Student Member,~IEEE}, Wenchi Cheng,~\IEEEmembership{Senior Member,~IEEE}, Zhanyu Ju,~\IEEEmembership{Student Member,~IEEE}, Jingqing Wang,~\IEEEmembership{Member,~IEEE}, Jinzhe Pan, Yuehui Ouyang and Wei Zhang, ~\IEEEmembership{Fellow,~IEEE} 
\thanks{Peini Yi, Wenchi Cheng, Zhanyu Ju and Jingqing Wang are with State Key Laboratory of Integrated Services Networks, Xidian University, Xian 710071, China (e-mails: pnyi@stu.xidian.edu.cn; wccheng@xidian.edu.cn; juzhanyu@stu.xidian.edu.cn; jqwangxd@xidian.edu.cn.)}
\thanks{Jinzhe Pan and Yuehui Ouyang are with the Wireless Communication Department, Honor Device Co. Ltd., Shenzhen, China (e-mails: panjinzhe@honor.com; yuehuiouyang@honor.com).}
\thanks{Wei Zhang is with the School of Electrical Engineering and Telecommunications, University of New South Wales, Sydney, NSW 2052, Australia
(e-mails: w.zhang@unsw.edu.au)}}



\maketitle
\thispagestyle{empty}
\pagestyle{empty}
\begin{abstract}

With the increasing complexity of Wi-Fi networks and the iterative evolution of 802.11 protocols, the Carrier Sense Multiple Access with Collision Avoidance (CSMA/CA) protocol faces significant challenges in achieving fair channel access and efficient resource allocation between legacy and modern Wi-Fi devices. To address these challenges, we propose an AI-driven Station (AI-STA) equipped with a Deep Q-Learning (DQN) module that dynamically adjusts its receive sensitivity threshold and transmit power. The AI-STA algorithm aims to maximize fairness in resource allocation while ensuring diverse Quality of Service (QoS) requirements are met. The performance of the AI-STA is evaluated through discrete event simulations in a Wi-Fi network, demonstrating that it outperforms traditional stations in fairness and QoS metrics. Although the AI-STA does not exhibit exceptionally superior performance, it holds significant potential for meeting QoS and fairness requirements with the inclusion of additional MAC parameters. The proposed AI-driven Sensitivity and Power algorithm offers a robust framework for optimizing sensitivity and power control in AI-STA devices within legacy Wi-Fi networks.

\end{abstract}


\section{Introduction}
\IEEEPARstart{S}{ince} its introduction in 1997, wireless communication networks based on the IEEE 802.11 standard have become a cornerstone of modern connectivity \cite{pahlavanWidebandLocalAccess1997}. At the core of the 802.11 Medium Access Control (MAC) layer is the Carrier Sense Multiple Access with Collision Avoidance (CSMA/CA) protocol \cite{bianchiPerformanceEvaluationEnhancement1996}, which facilitates channel access by enabling devices to sense the channel before transmission and avoid collisions. While CSMA/CA serves as a robust channel access mechanism, its limitations become increasingly evident as Wi-Fi networks scale, particularly in high-density deployments \cite{pyattaevCommunicationChallengesHighdensity2015,ganjiCharacterizingPerformanceWiFi2019} with diverse Quality of Service (QoS) \cite{mengIEEE80211Real2018,wangStatisticalDelayErrorRate2024} requirements. In such scenarios, heterogeneous devices face significant challenges, such as unequal channel access and inefficient resource allocation, further complicating network performance optimization.

To address these limitations, the concept of AI-driven Medium Access Control (AI-MAC) has emerged as a promising solution \cite{pan2024macrevivoartificialintelligence,zhangMultipleAccessIntegrated2024}. AI-MAC utilizes machine learning algorithms to dynamically adapt transmission strategies in real-time, introducing a flexible and intelligent framework for network resource management. This approach enables devices to optimize performance in complex, multi-device environments while ensuring fairness and maintaining QoS. The application of AI to MAC protocols holds the potential to revolutionize Wi-Fi network management, enhancing channel access, resource allocation, and QoS optimization \cite{szottWifiMeetsML2022}.

A critical aspect of Wi-Fi network performance involves the interplay between the Clear Channel Assessment (CCA) and Transmit Power Control (TPC) mechanisms through the optimization of spatial reuse \cite{xiaTransmitPowerControl2015,ramachandranClearChannelAssessment2007}.
These mechanisms are pivotal in managing channel utilization and controlling interference among devices. TPC governs the signal strength of transmissions, with higher TPC settings extending signal coverage and expanding the network's range. However, this advantage comes at the cost of increased interference, particularly in Overlapping Basic Service Set (OBSS) scenarios, where it may disrupt communication for other devices. On the other hand, the CCA mechanism determines a device's ability to detect signals from other devices. A lower CCA threshold enhances the device's sensitivity to distant signals, reducing potential transmission conflicts by avoiding data transmission when the channel is occupied. Conversely, a higher CCA threshold risks overlooking signals, leading to increased conflicts and degraded network performance.

In addition, the coexistence of Wi-Fi devices across multiple generations introduces a significant technical challenge \cite{dengIEEE80211beWiFi2020,kosek-szottCoexistenceIssuesFuture2017}. Ensuring compatibility among devices with varying capabilities is essential for harmonious operation within the 802.11 airspace, especially in unlicensed spectra. Backward compatibility not only ensures effective integration but also maintains the functionality of legacy devices, a critical consideration. It is crucial to ensure fair channel access for legacy devices while fulfilling their transmission needs. This also facilitates cooperation between new and legacy devices in the Wi-Fi network.

Significant efforts have been devoted to enhancing communication performance through CCA threshold adjustment strategies. For instance, the Dynamic Sensitivity Control (DSC) algorithm proposed in \cite{DynamicSensitivityControl} adjusts the CST of a user based on the Received Signal Strength Indicator (RSSI) from its associated AP. This approach aims to mitigate interference and improve fairness.
In \cite{lvAdaptiveRateCarrier2017,huaDistributedPhysicalCarrier2009}, fairness is quantified by incorporating the Packet Error Rate (PER) of other devices into a fairness ratio, which determines whether to adjust the CCA threshold. Channel occupancy is another factor considered for CCA threshold adjustment, as discussed in \cite{zhuOptimalPhysicalCarrier2007}.
Additionally, joint optimization of the CCA threshold and Transmit Power Control (TPC) has been widely explored to improve spatial reuse and overall system performance \cite{afaquiEvaluationDynamicSensitivity2015,kimFACTFineGrainedAdaptation2017,kimImprovingSpatialReuse2006,fuemmelerSelectingTransmitPowers2006}. Overall, these studies highlight the effectiveness of CCA threshold adjustment strategies in improving communication performance, while leaving opportunities to explore more adaptive and efficient approaches.

Deep Reinforcement Learning (DRL) has been widely applied in WiFi wireless networks to optimize various aspects such as channel allocation, rate adaptation, channel access, TPC, and sensitivity adjustment. For instance, Nakashima et al. in \cite{nakashimaDeepReinforcementLearningBased2020} proposed a DRL-based channel allocation scheme to maximize throughput in densely deployed multi-OSS wireless LANs. Similarly, Cho et al. in \cite{choReinforcementLearningRate2021} designed a reinforcement learning agent to control data transmission rates in CSMA/CA-based networks. Cakir et al. in \cite{cakirDTWNQlearningbasedTransmit2022} introduced a Digital Twin WiFi Network (DTWN) framework to reduce latency and improve system performance. Huang et al. in \cite{huangDeepQNetworkApproach2022,huangThreeTierDeepLearningBased2023} combined CCA and TPC optimization via DRL algorithms to enhance network capacity and minimize average queue length. While these studies collectively demonstrate the potential of DRL in optimizing various aspects of Wi-Fi networks, most of them overlook critical factors, including fairness between legacy and new devices and the QoS requirements of different services.

To overcome the abovementioned challenges, in this paper, we focus on optimizing the performance of a single AI-driven Station (\text{AI-STA}) within a legacy Wi-Fi network. The \text{AI-STA} is equipped with an artificial intelligence module that dynamically adjusts its receive sensitivity threshold and transmit power in response to real-time channel conditions and network dynamics. Our objective is to enhance the sensitivity, power control, and fairness of the \text{AI-STA} while satisfying diverse QoS requirements. To this end, we propose a novel DQN-based AI-STA algorithm based on Deep Q-Learning (DQN) to optimize these parameters. The proposed algorithm is designed to maximize fairness in resource allocation while maintaining QoS, thereby improving the overall performance of the \text{AI-STA} in legacy Wi-Fi networks.

The remainder of this paper is structured as follows: Section II describes the system model of the proposed \text{AI-STA} in a legacy Wi-Fi network, including the network model, the channel model, and the link model. Section III introduces the optimization objectives, including fairness metric and QoS performance metrics. Section IV presents the DQN-based AI-STA algorithm for optimizing the sensitivity and power of the \text{AI-STA}. Section V discusses simulation results and evaluates the performance of the proposed solution. Finally, Section VI concludes the paper.

\section{System Model}

In this section, we propose a model for a single AI-driven STA (\text{AI-STA}) in a legacy Wi-Fi network, as illustrated in Fig. 1. The \text{AI-STA} is equipped with an AI module that dynamically adjusts its receive sensitivity threshold and transmit power strategy based on real-time channel conditions and network dynamics. We will analyze the network model, channel model, and link model for the scenario under consideration.
\begin{figure}[t]
  \centering
  \includegraphics[width=3.4in]{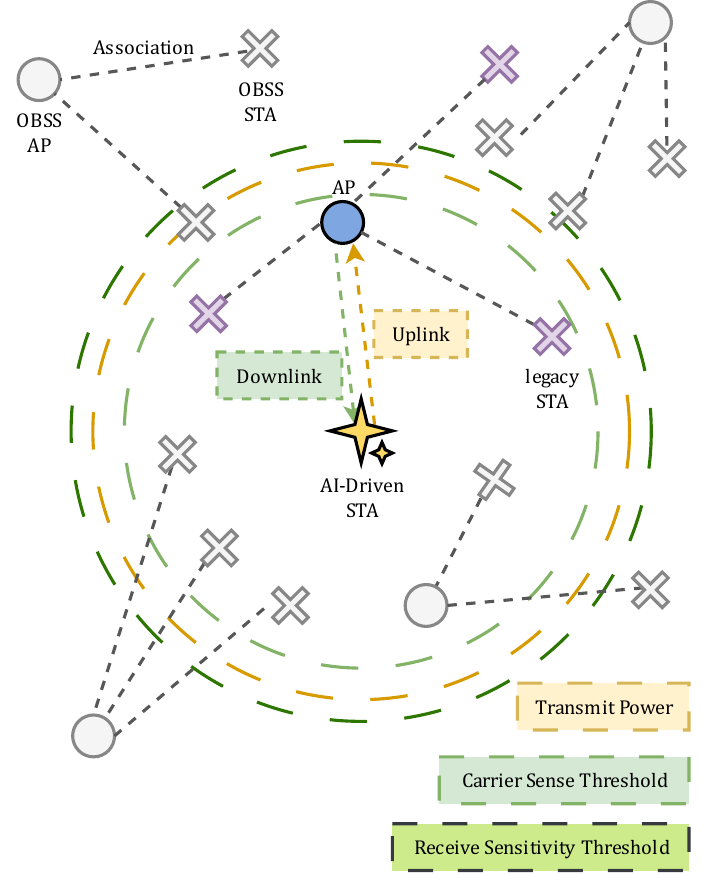}
  \caption{The single AI-driven STA in a legacy Wi-Fi network scenario}
  \label{fig:system_model}
\end{figure}

\subsection{Network Model}
In the wireless environment, the device set \(\mathcal{D}\) comprises \(\mathrm{n}\) Wi-Fi devices, consisting of \(\mathrm{m}\) access points (APs) and \(\mathrm{l}\) stations (STAs), i.e.,

\begin{equation}
\mathcal{D} = \mathcal{D}_{\text{AP}} \cup \mathcal{D}_\text{STA}
\end{equation}
where \(\mathcal{D}_{\text{AP}} = \{ AP_1, AP_2, \dots, AP_\mathrm{m} \}\) denotes the set of AP devices, and \(\mathcal{D}_{\text{STA}} = \{ \text{STA}_1,\text{STA}_2, \dots, \text{STA}_{\mathrm{l}-1},\text{STA}_\text{AI} \}\) denotes the set of STA devices. Among the devices in \(\mathcal{D}_{\text{STA}}\), \(\text{STA}_\text{AI}\) is the sole AI-driven STA device, and the remaining \(n-1\) devices are legacy. The set of legacy devices, \(\mathcal{D}_\text{legacy}\), is defined as:

\begin{equation}
\mathcal{D}_\text{legacy} = \mathcal{D}_\text{AP} \cup \{ \text{STA}_1, \text{STA}_2, \dots, \text{STA}_{n-1} \}
\end{equation}

In wireless networks, each STA device is associated with an AP device. We can represent the association between STA devices and AP devices by defining an association set \(a \subseteq \mathcal{D}_{\text{STA}} \times \mathcal{D}_{\text{AP}}\), where each element \((\text{STA}_i, \text{AP}_j)\) represents the association of STA device \(\text{STA}_i\) with AP device \(AP_j\). That is:
\begin{equation}
a = \{ (\text{STA}_i, \text{AP}_j) \mid \text{\text{STA}}_i \text{ is associated with } \text{\text{AP}}_j \}
\end{equation}

\subsection{Channel Model}
Each associated STA and AP device in the wireless environment will establish a communication link with one another and can function as transceivers for each other. Consider any pair of transceiver links, where the transmitter is denoted as \( s \) and the receiver as \( r \). The path loss between them can be modeled as:
\begin{equation}
PL(d) = \left( \frac{d}{d_0} \right)^\alpha \psi
\end{equation}
where \(d\) is the distance between the transmitter and receiver, \(\alpha\) is the path loss exponent, which depends on the environment type, with typical indoor path loss exponents ranging from 2 to 4, \(d_0\) is the reference distance, \(\psi\) is modeled as \( \mathcal{CN}(0, N_0) \) Gaussian white noise.

To more accurately describe the effects of multipath fading and shadowing in indoor environments, the distribution of all received signals follows the Nakagami-m distribution\cite{goldsmithWirelessCommunications}. The power gain distribution of the Nakagami-m fading channel is given by the following formula:
\begin{equation}
p_{|h|^2}(z)=\frac{m^m}{\Gamma(m)}z^{m-1}e^{-mz},\quad z\geq0
\end{equation}
where \(|h|^2\) is the received power following the Nakagami-\(m\) fading distribution, \(m\) is the shape factor of the Nakagami-\(m\) distribution, typically ranging from 1 to 2, and \(\Gamma(m)\) is the Gamma function.

Under the effects of path loss and Nakagami-m fading, the received signal power caused by transmitter \(P_{s,r}\) at the AP device can be expressed as:
\begin{equation}
 P_{s,r} = P_{s} \cdot PL(d) \cdot |h|^2
  \end{equation}
Here, \(P_{s}\) is the transmit power of the source, \(PL(d)\) is the path loss, and \(|h|^2\) is the power gain following the Nakagami-m fading distribution.

The interference on the receiver, denoted by \(I_{r}\), comes from all other transmitting nodes outside the transmitter. The interference power can be expressed as:
\begin{equation}
I_{r} = \sum_{x \in \mathcal{D},x\neq s } P_{x,r} = \sum_{\mu \in \mathcal{D},x\neq s } P_{s} \cdot PL(d) \cdot |h|^2
\end{equation}

When receiving from the source, the signal interference noise ratio (SINR) of the link between the receiver and transmitter can be expressed as:
\begin{equation}
\gamma = \frac{P_{s,r}}{{I_r} + N_0}
\end{equation}

\subsection{Link Model of \text{AI-STA} Model}
Our optimization target is the QoS performance of a single AI-driven STA within a BSS consisting of legacy devices. We aim to achieve this by changing the transmit power and receive sensitivity parameters of the AI-driven STA. Therefore, we will analyze the uplink and downlink of the AI-driven STA and its associated AP.

Here, we identify three key parameters: the Carrier Sensing Threshold (CST), the Receiver Sensitivity Threshold (RST), and the transmit power of the \text{AI-STA} (\(P_s\)). The CST represents the minimum power level required to detect that the channel is occupied, while the RST denotes the minimum power level at which the receiver can successfully decode the signal. The transmit power of the \text{AI-STA} corresponds to the power level at which the \text{AI-STA} transmits data to the AP. In the following, we analyze how these three parameters influence the transmission and reception processes of the AI-STA.

\subsubsection{Uplink} The \text{AI-STA} sends data to the AP. Here the AP is the receiving node and the STA is the sending node, our goal is to adjust the transmit power of the STA \(P_s\) because the CST of the AP is determined, so it affects whether the sent packets are received or not, and the transmit power of the AP should be:

\begin{equation}
 P_{\text{AI-STA}} = \frac{P_{\text{AI-STA,AP}}}{PL(d) \cdot |h|^2} 
\end{equation}
where the $P_{\text{AI-STA}}$ is the transmit power of the \text{AI-STA}, and the $P_{\text{AI-STA},AP}$ is the transmit power of the AP to the \text{AI-STA}.

The power received by the AP from the transmitting node should be above a given receiver sensitivity threshold(RST) in order to be decoded correctly.
\begin{equation}
P_{\text{AI-STA}} \geq \frac{\xi_{rs}(\text{AP})({I_{\text{AP}} + N_0})}{PL(d) \cdot |h|^2}
\label{eq:transmit_power}
\end{equation}
where the \(\xi_{rs}\) is the receiver sensitivity threshold of the AP, and the \(P_{\text{AI-STA}}\) is the transmit power of the \text{AI-STA}.

The equation (\ref{eq:transmit_power}) provides only a lower bound for the transmission power of the \text{AI-STA}. The actual transmission power of the STA must exceed this lower bound to ensure that other devices can detect the \text{AI-STA} and avoid collisions with it. 

When the interference power is larger than the carrier sensitivity threshold, the \text{AI-STA} will choose not to send data to the AP, and the equation can be expressed as:

\begin{equation}
 I_{\text{AI-STA}} = \sum_{x \in \mathcal{D},x\neq \text{AI-STA} } P_{x,r} \geq \xi_{cs}(\text{AI-STA})
  \label{eq:interference}
\end{equation}
where the \(\xi_{cs}(\text{AI-STA})\) is the carrier sensing threshold of the STA, and the \(I_{\text{AI-STA}}\) is the interference power of the STA.

From this perspective, it may seem that increasing the \(P_s\) of a device would be beneficial, assuming no energy constraints. However, a higher \(P_s\) results in increased interference with the communication of other devices, potentially degrading the overall performance of the network. Therefore, the optimization of the transmit power of the \text{AI-STA} is a trade-off between ensuring that the \text{AI-STA} can be detected by other devices and minimizing interference with other devices.

\subsubsection{Downlink} The AP sends data to the \text{AI-STA}. In the downlink of this link, the AP is the receiving node and the STA is the transmitting node, and we are concerned with the receive sensitivity threshold, and the link's SINR \(\gamma\) can be expressed as:
\begin{equation}
{\xi}_{rs}(\text{AI-STA}) \leq {\frac{P_{\text{AP,AI-STA}}}{I_{\text{AI-STA}} + N_0}} = \frac{P_{\text{AP}} \cdot PL(d) \cdot |h|^2}{I_{\text{AI-STA}} + N_0}
\end{equation}
where the $P_{\text{AP},\text{AI-STA}}$ is the transmit power of the AP to the \text{AI-STA}, and the $RST_{STA}$ is the receiver sensitivity threshold of the STA.

In fact, the relationships among transmit power, CST, and RST are complex and interdependent. To improve channel competition in varying environments, the AI-STA dynamically adapts its CST to meet its QoS requirements. Surrounded by legacy devices that do not adapt to the environment, the AI-STA must adjust both its transmit power and RST to ensure reliable communication with the associated AP in both uplink and downlink directions while minimizing interference with other devices. This optimization problem is particularly challenging due to the dynamic and stochastic nature of the wireless environment. In the following sections, we will introduce the optimization objectives and the DQN-based AI-STA algorithm designed to tackle this challenge.

\section{Optimization Target}
\label{sec:optimization_target}
In this section, we will introduce the optimization target of the \text{AI-STA} in the legacy Wi-Fi network. We will analyze the fairness metrics and QoS performance metrics of the \text{AI-STA}, and formulate the optimization problem.
In this section, we will introduce the optimization target of the \text{AI-STA} in the legacy Wi-Fi network. We will analyze the fairness metrics and QoS performance metrics of the \text{AI-STA}, and formulate the optimization problem.

\subsection{Independent Fairness Metric}

\begin{figure*}[t]
  \centering
  \includegraphics[width=7in]{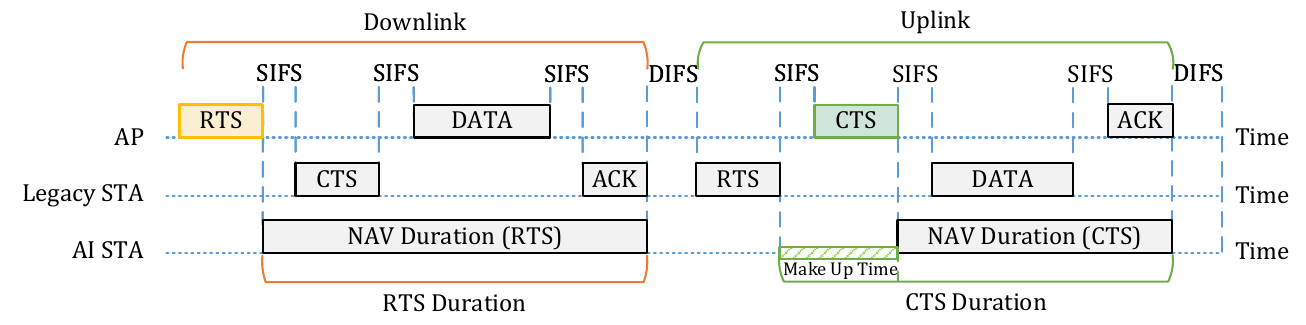}
  \caption{The \text{AI-STA} Listening to RTS/CTS Packets and Updating STA-Duration Table}
  \label{fig:fair_detail}
\end{figure*}

Analyzing access fairness is crucial in wireless networks to ensure equitable resource allocation and maintain Quality of Service. While achieving better QoS, higher throughput, lower latency, and smaller jitter may suggest adopting an aggressive channel strategy, such as using the highest Clear Channel Assessment threshold for detecting AP nodes, this approach can lead to excessive channel contention \cite{cagaljSelfishBehaviorCSMA2005}. Such aggressive contention would restrict other devices within the same Basic Service Set from accessing the channel, thereby degrading their QoS.

In contrast to traditional access points, the AI-driven STA lacks comprehensive uplink and downlink data statistics for the entire BSS. This limitation makes it challenging for the STA to assess overall channel fairness and recognize its own potential for excessive channel contention. Therefore, it is necessary to introduce a penalty mechanism for uncivilized channel contention, which encourages the \text{AI-STA} to self-reflect and adjust its channel behavior.

Although several studies about fairness have analyzed fairness in BSS environments, most of them focus on AP nodes, leaving the assessment of fairness from the STA perspective largely unexplored. The inability of STAs to access global uplink and downlink information within the BSS further complicates fairness assessment.

To address this challenge, this paper proposes a framework for independent access fairness based on control frames in IEEE 802.11 protocol \cite{IEEEStandardInformation2011}. This framework enables STA nodes within a BSS to independently evaluate fairness while considering the actions of other devices in the same BSS. 

The \text{AI-STA} will maintain a table that records time cost in the BSS based on the network allocation vector (NAV). In detail, the MAC addresses (as key values) of other STA nodes detected within the same BSS, along with the total duration these nodes occupy the channel. This table will be updated in real-time by the \text{AI-STA} based on the duration specified in all RTS/CTS packets sent by the monitored AP, along with the corresponding destination node address.

When the \text{AI-STA} updates the duration table, it will also consider the time difference for RTS duration and CTS duration. As shown in Fig. \ref{fig:fair_detail}, to equalize the time difference between the RTS and CTS packets, the \text{AI-STA} will add the CTS time and SIFS time to the CTS duration. The proposed duration update algorithm is outlined as given in Algorithm \ref{alg:process_rts_cts}.

After compiling the results from the table, the \text{AI-STA} evaluates the fairness of channel time usage, denoted as \(F\), using Jain's fairness coefficient. This coefficient quantifies the degree of equilibrium in channel occupation time among all STAs within the shared channel.

\begin{algorithm}[H]
  \caption{Process RTS/CTS Packets and Update STA-Duration Table}\label{alg:process_rts_cts}
  \begin{algorithmic}
  \FORALL{packet $\in$ RTS or CTS packets from associated AP}
      \IF{packet type is RTS}
      \STATE key $\leftarrow$ get key from receive address\;
          \IF{key exists in STA-duration table}
          \STATE duration[key] $\leftarrow$ duration[key] + packet duration\;
          \ENDIF
      \ELSIF{packet type is CTS}
          \STATE key $\leftarrow$ get key from receive address\;
          \IF{key exists in STA-duration table}
              \STATE duration[key] $\leftarrow$ duration[key] + CTS + SIFS + packet duration\;
          \ENDIF
      \ENDIF
  \ENDFOR
  \end{algorithmic}
\end{algorithm}

Let \(T_i\) denote the channel occupation time of the \(i\)-th STA, and let \(N\) represent the total number of STAs within the BSS. Jain's fairness coefficient \(F\) is expressed as:

\begin{equation}
 F = \frac{\left(\sum_{i=1}^{N} T_i\right)^2}{N \cdot \sum_{i=1}^{N} T_i^2}
  \label{eq:fair}
\end{equation}

\(F = 1\) indicates perfect fairness, where all STAs equally share the channel. Conversely, \(F < 1\) signifies the presence of unfairness in channel usage.

\subsection{QoS Performance}

To meet diverse QoS requirements, the performance of the AI-driven device is evaluated based on key metrics, including packet loss rate, throughput, delay, and jitter. Given the complexity of network environments, deriving a closed-form expression for the QoS performance of the \text{AI-STA} is challenging. Consequently, we utilize discrete event simulation to assess and analyze the QoS performance of the \text{AI-STA} in a Wi-Fi network. Below, we introduce the key QoS metrics used in the simulation.

\subsubsection{Throughput}

Throughput is defined as the amount of data transmitted over a network within a given period of time. It is a critical metric for assessing the QoS of a network, particularly in services such as video streaming. The throughput, denoted as \(\bar{S}\), is calculated as the total amount of data transmitted over a specified period. The throughput is given by:

\begin{equation}
  \bar{S} = \frac{\sum_{i=1}^{N} \text{Data}_i}{T(\text{End}) - T(\text{Start})}
  \label{eq:throughput}
\end{equation}
where \(\text{Data}_i\) is the size of the \(i\)-th packet. Throughput is measured in bits per second (bps). A higher throughput indicates better QoS performance, as it signifies the ability to transmit more data over the network in a given period of time.

\subsubsection{Latency}

The delay of a packet is defined as the time taken for the packet to travel from the source to the destination. The packet delay is computed as the difference between the time the packet is sent and the time it is received at the destination. The delay of a packet is given by:

\begin{equation}
 D = T(\text{Packet Receive}) - T(\text{Packet Send})
\end{equation}
where \(T(\cdot)\) refers to the timer in the discrete event simulation.

The packet delay is measured in milliseconds (ms). A lower delay indicates better QoS performance, as it reflects quicker delivery of packets from the source to the destination.

In this context, we use latency, represented by the average packet delay \(\bar{D}\), as the metric to evaluate the QoS performance of the \text{AI-STA}. The average delay is calculated as the mean delay of all packets sent by the \text{AI-STA}. A lower average delay indicates better QoS performance. The average delay is computed as follows:

\begin{equation}
  \bar{D} = \frac{\sum_{i=1}^{M} D_i}{M}
  \label{eq:average_delay}
\end{equation}
where \(M\) is the total number of packets sent by the \text{AI-STA}, and \(D_i\) represents the delay of the \(i\)-th packet.

\subsubsection{Jitter}

Jitter is defined as the variation in the delay of packets. Jitter is calculated as the difference between the delays of two consecutive packets. The average jitter is computed as follows:

\begin{equation}
  \bar{J} = \frac{1}{N-1} \sum_{i=1}^{N-1} |D_{i+1} - D_i|
  \label{eq:average_jitter}
\end{equation}
where \(N\) is the total number of packets, and \(D_i\) represents the delay of the \(i\)-th packet.

\subsubsection{Packet Loss Rate}

The packet loss rate is defined as the percentage of packets lost during transmission. The packet loss rate, denoted by \(p_\text{loss}\), is calculated as the ratio of lost packets to the total number of packets sent. The packet loss rate is given by:

\begin{equation}
 p_\text{loss} = \frac{N - N_\text{recv}}{N}
  \label{eq:packet_loss_rate}
\end{equation}
where \(N_\text{recv}\) is the total number of packets received by the associated AP.

The packet loss rate is measured as a percentage. A lower packet loss rate indicates better QoS performance, as it reflects fewer lost packets during transmission.

\subsection{Problem Formulation}

With the QoS Performance and Fairness metrics mentioned above, let's analyze the optimization goals.

Different services have different QoS requirements, and they usually have minimum requirements for individual performance parameters, fairness is the part where we choose the best and keep them relatively well. STA needs to adjust the sensitivity threshold and transmit power to meet the QoS requirements of the service and make time for nodes within the same BSS to use the channel more fairly.

So our optimization problem can be expressed as follows:
\begin{equation}
  \begin{aligned}
\mathbf{P1}: \quad 
      \text{max} \quad 
      & F(\xi_{cs}, \xi_{rs}, P_{s}), \\
      \text{s.t.} \quad
      & S(\xi_{cs}, \xi_{rs}, P_{s}) \geq S^{\text{min}}, \\
      & D(\xi_{cs}, \xi_{rs}, P_{s}) \leq D^{\text{max}}, \\
      & J(\xi_{cs}, \xi_{rs}, P_{s}) \leq J^{\text{max}}, \\
      & p_{\text{loss}}(\xi_{cs}, \xi_{rs}, P_{s}) \leq p_{\text{loss}}^\text{max}, \\
      & \xi_{cs} \in [\xi_{cs}^{\text{min}}, \xi_{cs}^{\text{max}}], \\
      & \xi_{rs}\in [\xi_{rs}^{\text{min}}, \xi_{rs}^{\text{max}}], \\
      & P_{s} \in [P_s^{\text{min}}, P_s^{\text{max}}].
  \end{aligned}
  \label{eq:optimization_problem}
  \end{equation}
 where \(F(\cdot)\), \(S(\cdot)\), \(D(\cdot)\), \(J(\cdot)\), and \(p_{\text{loss}}(\cdot)\) represent the fairness, throughput, delay, jitter, and packet loss rate, respectively. These metrics are functions of \((\xi_{cs}, \xi_{rs}, P_s)\). Additionally, \(S^{\text{min}}\), \(D^{\text{max}}\), \(J^{\text{max}}\), and \(p_{\text{loss}}^\text{max}\) denote the minimum throughput, maximum delay, maximum jitter, and maximum packet loss rate, respectively. The parameters \(\xi_{cs}^{\text{min}}\), \(\xi_{cs}^{\text{max}}\), \(\xi_{rs}^{\text{min}}\), \(\xi_{rs}^{\text{max}}\), \(P_s^{\text{min}}\), and \(P_s^{\text{max}}\) specify the lower and upper limits of the carrier sensitivity threshold, receiver sensitivity threshold, and transmit power, respectively.

Notice that the QoS restrictions won't be achieved due to different channel characteristics in different network environments. Since our optimization algorithm is based on large amounts of simulation data generated by discrete-event simulations and uses statistical methods to evaluate overall system performance, in the following chapters we will use a data-driven approach such as deep reinforcement learning to improve performance.

\section{DQN-based AI-MAC Algorithm}
In this section, we will provide a detailed introduction to how to use deep reinforcement learning (DRL) technology to optimize the fairness of resource allocation while meeting QoS requirements. Our goal is to find the optimal values of sensitivity and power in a given environment.

\subsection{Markov Decision Process}
In this section, we will outline how to transform the optimization equation we have established into a Markov Decision Process (MDP). 

Generally, such stochastic dynamic systems can be described using a Markov Decision Process. An MDP is defined as a quadruple \((\mathcal{S}, \mathcal{A}, r, \mathcal{P})\), where \(\mathcal{S}\) is the set of state spaces, \(\mathcal{A}\) is the set of action spaces, \(r\) is the immediate reward function, \(\mathcal{P}\) is the probability mapping of transitioning from the current state to a new state after taking an action. We will discuss them in the following sections.

\subsubsection{State Space}

The system is evaluated based on several key statistical parameters, including fairness, throughput, latency, jitter, and packet loss. Together, these parameters reflect the sensitivity and perception of the \text{AI-STA} to the interaction of its transmission services with the surrounding environment. These parameters constitute a state space that effectively describes the current state of the agent under consideration. The state space can be represented as:
\begin{equation}
  \mathcal{S}=\left\{F,S,D,J,p_{loss}\right\}
\end{equation}
These parameters are previously defined in (\ref{eq:fair}), (\ref{eq:throughput}), (\ref{eq:average_delay}), and (\ref{eq:average_jitter}), and used to evaluate the performance of the \text{AI-STA} in the current state. The state space \(\mathcal{S}\) is continuous and can be discretized into a finite set of states.

\subsubsection{Action Space}

Here, our action space is composed of three dimensions: the CST denoted by \(\xi_{cs}\), the RST denoted by \(\xi_{rs}\), and the transmit power denoted by \(P_s\). All three dimensions form continuous action spaces, which can be represented as:
\begin{equation}
  \mathcal{A} = \left\{ (\xi_{cs}, \xi_{rs}, P_s) \mid 
\begin{aligned}
\xi_{cs} &\in [T_{c}^{\text{min}}, \xi_{cs}^{\text{max}}],\\ 
\xi_{rs} &\in [\xi_{rs}^{\text{min}}, \xi_{rs}^{\text{max}}],\\
P_s &\in [P_s^{\text{min}}, P_s^{\text{max}}]
\end{aligned}
\right\}
\end{equation}
where \(\xi_{cs}^{\text{min}}\), \(\xi_{cs}^{\text{max}}\), \(\xi_{rs}^{\text{min}}\), \(\xi_{rs}^{\text{max}}\), \(P_s^{\text{min}}\), and \(P_s^{\text{max}}\) represent the lower and upper limits of CST, RST, and transmit power, respectively.

The units of these three action parameters are all in dBm, and they form the action granularity within the dynamic range with the same steps, denoted by $l$. Consequently, the action space can be further discretized as:

\begin{equation}
  \mathcal{A} = \left\{ (\xi_{cs}, \xi_{rs}, P_s) \mid 
\begin{aligned}
\xi_{cs} &\in \{ \xi_{cs}^{\text{min}} + i\Delta_{cs} \}_{i=0}^{l}, \\
\xi_{rs} &\in \{ \xi_{rs}^{\text{min}} + j\Delta_{rs} \}_{j=0}^{l}, \\
P_s &\in \{ P_s^{\text{min}} + k\Delta_{s} \}_{k=0}^{l} 
\end{aligned}
\right\}
\end{equation}
where, $i$, $j$ and $k$ represent the index of the action space, and \(\Delta_{(\cdot)} = \frac{1}{l}[{(\cdot)^{max}-(\cdot)^{min}}]\) represents the action granularity for each dimension. With the action space $  \mathcal{A}$, we can obtain the state of the MDP at iteration $t$ is $a_t = (\xi_{cs}, \xi_{rs}, P_s) \in \mathcal{A}$.

\subsubsection{Reward Function}

The optimization goal is to maximize fairness under the condition of satisfying QoS, we can convert the previously proposed optimization problem (\ref{eq:optimization_problem}) into a reward function by changing the constraints, and we can convert the statistical constraints into penalty terms defined as follows:
\begin{equation}
\begin{aligned}
\mathcal{\zeta}_{S} &=\max(0,S_{\text{min}}-S(a_t))\\ 
\mathcal{\zeta}_{D} &=\max(0,D(a_t)-D_{\text{max}})\\
\mathcal{\zeta}_{J} &=\max(0,J(a_t)-J_{\text{max}})\\
\mathcal{\zeta}_{p_\text{loss}} &=\max(0,p_{\text{loss}}(a_t)-p_{\text{loss}}^\text{max})
\end{aligned}
\end{equation}
where the $\zeta_{\cdot}$ represents the penalty term for the corresponding QoS metric, and the $\omega_{\cdot}$ represents the weight of the corresponding QoS metric in the reward function. At step $t$, the reward function $r(s_t, a_t)$ for taking action $a_t$ in state $s_t$ can be expressed as:
\begin{equation}
\begin{aligned}
r(s_t,a_t) 
= &F(a_t) \\
&+ \omega_{S}\zeta_{S} + \omega_\text{D} \zeta_{\text{D}} + \omega_{J}\zeta_{J} + \omega_{p_{\text{loss}}}\mathcal{\zeta}_{p_\text{loss}}
\end{aligned}
\end{equation}
where \(\omega_{S}\), \(\omega_{D}\), \(\omega_{J}\), and \(\omega_{p_{\text{loss}}}\) are negative constants represent the weights of the throughput, delay, jitter, and packet loss rate, respectively. These weights are used to adjust the relative importance of each QoS metric in the reward function.

\subsubsection{Transition Probability}

The state transition probability in a Markov Decision Process (MDP) defines the likelihood of transitioning from a current state \(s_t\) to a next state \(s_{t+1}\), given an action \(a_t\), denoted by \(\mathcal{P}(s_{t+1} | s_t, a_t)\). 

In the context of our system, this probability reflects the dynamics of the wireless environment, which are influenced by the chosen action parameters \(\xi_{cs}\), \(\xi_{rs}\), and \(P_s\), as well as external factors such as channel interference and traffic loads. In such a complex environment the state transition probability can not be delivered exactly. However, we can use a model-free reinforcement learning algorithm to approximate the transition probability. In the following sections, we will introduce the deep Q-Network (DQN) algorithm and how to use it to optimize the \text{AI-STA}'s sensitivity and power parameters.

\subsection{Deep Q-Network Algorithm}

In this section, we introduced the DQN algorithm to optimize the \text{AI-STA}'s sensitivity and power parameters. 

\subsubsection{Q-Function}

The Q-function, or action-value function, is a key component in reinforcement learning algorithms, particularly in Deep Q-Networks (DQN). It represents the expected cumulative reward of taking an action \(a_t\) in state \(s_t\) and following the optimal policy thereafter. The objective of DQN is to maximize the long-term reward by selecting the action with the highest Q-value.

To update the Q-values, DQN utilizes the temporal difference (TD) target\cite{zhao2024RLBook}, denoted as \(y_T\), which is computed as follows:
\begin{equation}
 y_T(s_t, a_t) = r_{t+1} + \gamma \max_{a \in \mathcal{A}(s_{t+1})} q_t(s_{t+1}, a) 
\label{eq:td_target}
\end{equation}
where, \(y_T(s_t, a_t)\) represents the current estimate of the Q-value for state \(s_t\) and action \(a_t\), while \(\gamma\) is the discount factor. A larger \(\gamma\) places more emphasis on future rewards, encouraging long-term planning, while a smaller \(\gamma\) prioritizes immediate rewards.

The Q-function is updated iteratively using the TD target, as shown by:
\begin{equation}
 q_{t+1}(s_t, a_t) = q_t(s_t, a_t) + \alpha(s_t, a_t) (y_t - q_t(s_t, a_t))
\end{equation}
where, \(\alpha(s_t, a_t)\) is the learning rate for $(s_t, a_t)$, determining how much the Q-value is adjusted in response to the new information. By repeatedly updating the Q-values in this manner, DQN converges towards the optimal action-value function, enabling the agent to make decisions that maximize long-term rewards.

\subsubsection{Network Update and Experience Replay}

DQN employs two neural networks to approximate the Q-function, the main and target networks. The main network, parameterized by \(w\), estimates the Q-values, while the target network, parameterized by \(w_T\), provides stable target Q-values. The TD target denoted by \(y_T(s_t, a_t)\) is computed by (\ref{eq:td_target}).
The main network is updated by minimizing the mean squared error (MSE) between the predicted Q-values and the TD targets:
\begin{equation}
\mathcal{L} = \frac{1}{N} \sum_{i=1}^N \left( q(s_i, a_i; w) - y_T(s_i, a_i) \right)^2
\end{equation}
The target network synchronizes with the main network periodically to stabilize learning.

To further enhance training efficiency, experience replay stores agent experiences \((s_t, a_t, r_t, s_{t+1})\) in a buffer \(\mathcal{B}\). During each training step, a mini-batch of samples is randomly drawn from \(\mathcal{B}\), breaking the temporal correlation and ensuring diverse training data. This combination of network separation and experience replay improves convergence and stability.

\subsubsection{Action Policy}
As our proposed MDP has a continuous action space, and the environment is complex, we adopt the on-policy learning to train the sensitivity and power parameters of \text{AI-STA}. The \( \varepsilon \)-greedy action policy is utilized to select the agent behavior, where \( \varepsilon \) is the exploration probability. Specifically, we select the optimal action of the policy \( \pi^* \) with probability \( \varepsilon \) and a random action with probability \( 1 - \varepsilon \). Finally, as the last step, the internal neurons can be updated using backpropagation.

\begin{algorithm}[H]
\caption{DQN Scheme for Optimize Sensitivity and Power}
\label{alg:DQN}
\begin{algorithmic}
\STATE Initialize the experience replay buffer $\mathcal{B}$, the main network $q(s, a; w)$ and the target network $q(s, a; w_T)$ with random weights $w$ and $w_T = w$, respectively
\FOR{each iteration}
\STATE Observe the current state $s_t \in \mathcal{S}$
\STATE Choose an action $a_t$ based on $\varepsilon$-greedy policy
\STATE Execute the action $a_t$ 
\STATE Observe the reward $r_t$ and the next state $s_{t+1}$
\STATE Store the transition $(s_t, a_t, r_t, s_{t+1})$ in the experience replay buffer $\mathcal{B}$
\STATE Sample a mini-batch of transitions $(s_i, a_i, r_i, s_{i+1})$ from $\mathcal{B}$
    \IF {Every 100 iteration}
    \STATE Copy the main network weights to the target network: $w_T = w$
    \ENDIF
\ENDFOR
\end{algorithmic}
\end{algorithm}

\section{Simulation Results}
In this section, we will introduce our implementation of the DQN-based AI-MAC algorithm and the simulation results.

\subsection{Simulation Setup}

The simulation is built upon ns-3, a discrete-event network simulator, along with its associated packages ns3-ai \cite{yinNs3aiFosteringArtificial2020} and ns3-vr-app \cite{lecciNs3ImplementationBursty2021}. The simulation settings are outlined in Table I. Notably, the target QoS settings are derived from \cite{mengIEEE80211Real2018} and are designed to meet the demands of real-time mobile gaming scenarios. The node topology, as shown in Fig. \ref{fig:system_model}, reflects a typical home Wi-Fi network environment, and the protocol is set to IEEE 802.11ax.

\begin{table}[h]
  \begin{center}
    \caption{Parameters for NS3 Simulation}
    \begin{tabular}{c|c|l} 
      \hline
      \textbf{Parameter} & \textbf{Value} & \textbf{Definition}\\
      \hline
 m & 5 & Number of total AP nodes\\
 l & 14 & Number of total STA nodes\\
      $\alpha$ & 3 & Path loss exponent\\
      $N_0$ & -100 dBm & Power of Gaussian white noise\\
      \hline
      $T_s$       & $ \text{16 / [10, 30]}$ dBm & Default transmit power and range \\
      ${\xi}_{rs}$ & $\text{-101 / [-110, -30]}$ dBm & Default RST and range \\
      ${\xi}_{cs}$ & $\text{-82 / [-100, -20]}$ dBm & Default CST and range \\
      \hline      
      $\bar S$ & 1.5 Mbps & Minimum average throughput \\
      $\bar D$ & 5 ms & Maximum average delay \\
      $\bar J$ & 2 ms & Maximum jitter \\
      $p_{\text{loss}}$ & $0.1 \%$ & Maximum packet loss rate \\
      \hline
    \end{tabular}
  \end{center}
\end{table}

Interference plays a critical role in this simulation. In this study, we examine a dense network scenario. The AI-STA manages both uplink and downlink streams, which are reconstructed from real-time game data. In the case of BSS interference, the AP transmits a video stream to each STA associated with it. In the case of OBSS interference, each AP randomly selects a single STA to receive a video stream. Legacy STAs continuously receive video streams from the AP during BSS interference.
The simulation runs for 500 time steps, with each step lasting 0.1 seconds. The learning rate is set to 0.001, the discount factor to 0.8, and the exploration probability to 0.1. The QoS metric weights are set as follows: $\omega_{S} = 1$, $\omega_{D} = 10$, $\omega_{J} = 10$, and $\omega_{p_{\text{loss}}} = 0.1$. The action granularity is set to $l = 1$, and the experience replay buffer size is $5000$ with a batch size of $32$.
The DQN agent is implemented using Python 3.12.3 and PyTorch 2.5.1, running on an Intel Xeon(R) Gold 6430 processor and an RTX 4090 GPU. The simulation results are presented in the following sections. For this simulation, we utilized only instant learning to better reflect real-world scenarios.

\subsection{Simulation Results}
In this section, we compare our DQN-based AI-STA algorithm with a Dynamic Sensitivity Control (DSC) algorithm \cite{afaquiEvaluationDynamicSensitivity2015} and the traditional Wi-Fi system based on 802.11ax as the baseline algorithm.  

\begin{figure}[t]
  \centering
  \includegraphics[width=3.4in]{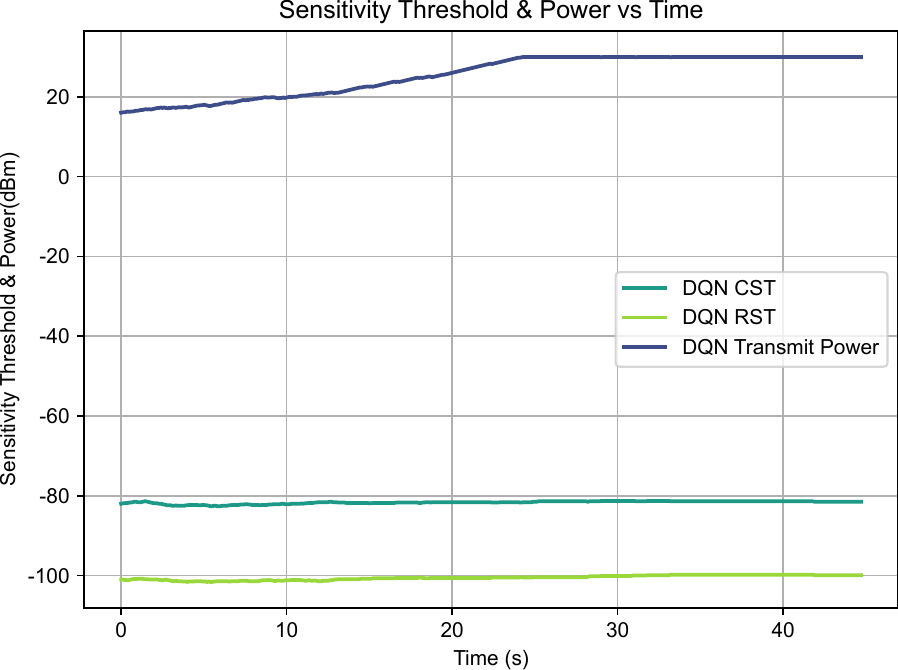}
  \caption{The DQN with varying of time}
  \label{fig:sensing}
\end{figure}

Fig. \ref{fig:sensing} shows the evolution of CST, RST, and transmit power over time for the DQN-based AI-STA algorithm. The algorithm dynamically adjusts the CST, RST, and transmit power. Over time, the CST and RST converge to their optimal values, while the transmit power stabilizes. This result demonstrates that the DQN-based AI-STA algorithm effectively adapts to the environment and optimizes sensitivity and power parameters to achieve optimal performance.

\begin{figure}[t]
  \centering
  \includegraphics[width=3.4in]{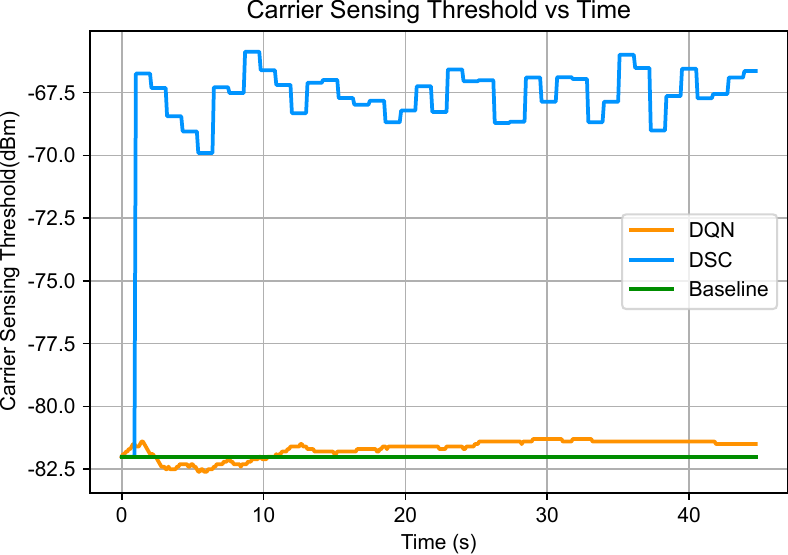}
  \caption{CST comparison of DQN-based, DSC, and baseline algorithm varying of time.}
  \label{fig:cst}
\end{figure}

In Fig. \ref{fig:cst}, we compare the CST behavior of the DQN-based AI-STA algorithm, the DSC algorithm, and the baseline algorithm. Unlike the traditional Wi-Fi system, both the DQN-based AI-STA and DSC algorithms dynamically adjust the CST to adapt to varying environmental conditions. A higher CST leads to more collisions, making CST an indicator of the algorithm's aggressiveness. Our results show that, compared to the DSC algorithm, the DQN-based AI-STA algorithm is less aggressive and closer to the baseline algorithm.

\begin{figure}[t]
  \centering
  \includegraphics[width=3.4in]{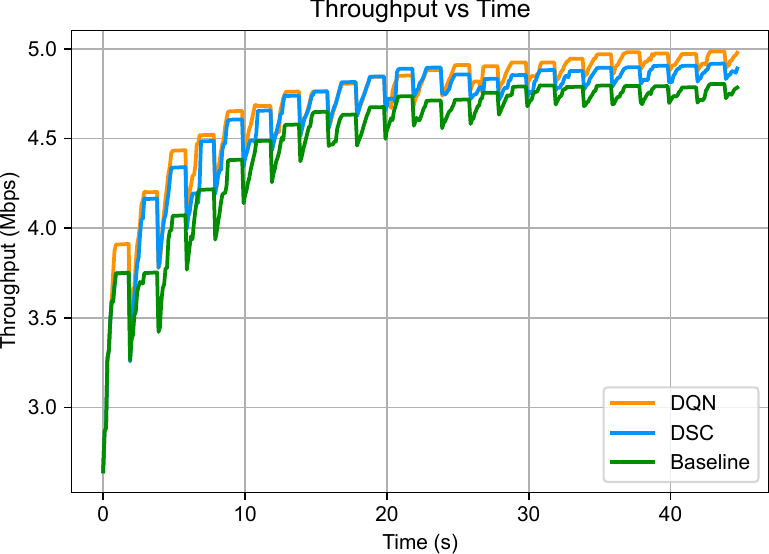}
  \caption{Throughput comparison of DQN-based, DSC, and baseline algorithm varying of time.}
  \label{fig:Throughput}
\end{figure}

Fig. \ref{fig:Throughput} illustrates the throughput comparison of the DQN-based AI-STA algorithm, the DSC algorithm, and the baseline algorithm over time. The results show that the throughput of the DQN-based AI-STA algorithm consistently outperforms both the DSC and baseline algorithms. However, under this competitive scenario, channel fading remains the primary factor influencing throughput performance, limiting further improvements. Additionally, the throughput graph reveals a periodic burst pattern in the data. Over time, nodes accumulate packets requiring retransmission, leading to a smoother throughput curve. This smoothing effect, however, contributes to increased jitter and delay.

\begin{figure}[t]
  \centering
  \includegraphics[width=3.4in]{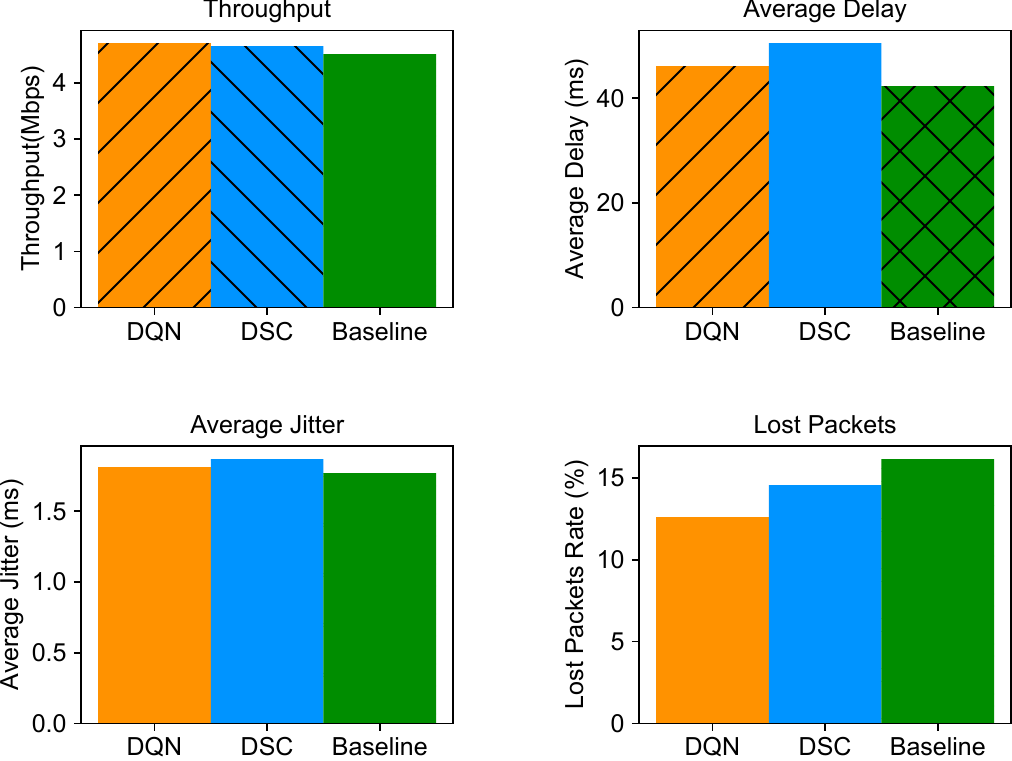}
  \caption{Average QoS metrics comparison of DQN-based, DSC, and baseline algorithm.}
  \label{fig:QoS}
\end{figure}

Fig. \ref{fig:QoS} compares all of the QoS metrics of the DQN-based AI-STA algorithm with the traditional Wi-Fi system and the DSC algorithm. The DQN-based AI-STA algorithm achieves higher throughput and lower packet loss rate compared to the baseline algorithm and the DSC algorithm, but its average delay and jitter are higher than the baseline algorithm. The poor performance in delay and jitter is primarily due to the fact that our algorithm only adjusts CST, RST, and transmission power, which are only part of the 802.11 MAC layer. While these parameters are used to optimize the QoS metrics and fairness metric, they do not account for all the factors that influence these metrics, particularly in crowded channel conditions. In such scenarios, parameters like contention window, backoff time, and retry limit have a significant impact on delay and jitter. More importantly, these parameters are closely related to the CST, RST, and transmission power that we adjust. For example, increasing CST inevitably increases the collision probability, which, in turn, affects performance. Therefore, our algorithm has certain limitations. Incorporating additional MAC parameters into the training process could potentially yield results that are more aligned with QoS requirements.

\begin{figure}[t]
  \centering
  \includegraphics[width=3.4in]{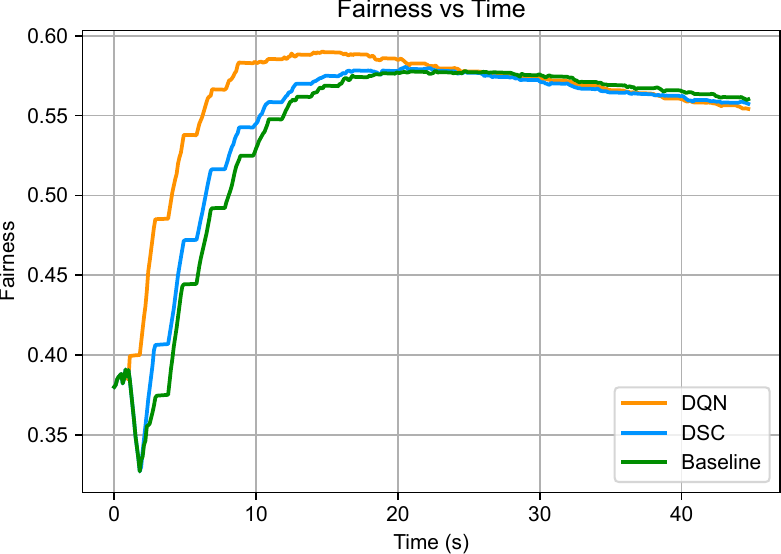}
  \caption{Fainess comparison of DQN-based, DSC, and baseline algorithm varying of time.}
  \label{fig:Fairness}
\end{figure}

Fig. \ref{fig:Fairness} illustrates the fairness metrics of the DQN-based AI-STA algorithm, the baseline algorithm, and the DSC algorithm over time. Fairness is our main optimization target, and the DQN-based AI-STA algorithm achieves a higher fairness score compared to the traditional Wi-Fi system and the DSC algorithm. From the overall fairness trend, it is evident that the DQN algorithm achieves significant fairness improvements during the early stages. However, as the simulation progresses, the impact of the discount factor reduces the rewards for actions taken in later stages. This results in a decrease in the algorithm's action frequency compared to the early stages. Over time, the communication pattern shifts from burst communication to saturation communication. As a result, the fairness index of the DQN algorithm gradually converges to the level of the DSC and baseline algorithms and may even slightly fall below the baseline in some cases.

\section{Conclution}

In this paper, we propose an AI-STA algorithm based on DQN, which aims to strike a balance between AI-STA devices and traditional devices while ensuring QoS. By leveraging the relationship between CCA and TPC, we ensure QoS in complex channel environments and provide fair transmission opportunities for traditional devices. Additionally, we develop a simulation based on ns-3 to explore the performance of this approach.
Our comprehensive analysis and simulation results demonstrate that DQN can optimize fairness by learning the channel environment and interference conditions, and adapting CCA and TPC strategies while maintaining optimal QoS. The results indicate that with sufficient learning, the DQN-based AI-STA algorithm converges to the optimal value. Compared to traditional algorithms, our proposed approach strikes a better balance between QoS and fairness. However, relying solely on sensitivity and power management is not sufficient to fully optimize AI-device performance, more adjustable MAC parameters are needed.
In summary, this study highlights the potential of DQN in optimizing Wi-Fi performance with fairness requirements and provides new insights into the design of innovative devices.


\bibliographystyle{IEEEtran}
\bibliography{JSAC1130AIMAC.bib}

\begin{thebibliography}{10}
\providecommand{\url}[1]{#1}
\csname url@samestyle\endcsname
\providecommand{\newblock}{\relax}
\providecommand{\bibinfo}[2]{#2}
\providecommand{\BIBentrySTDinterwordspacing}{\spaceskip=0pt\relax}
\providecommand{\BIBentryALTinterwordstretchfactor}{4}
\providecommand{\BIBentryALTinterwordspacing}{\spaceskip=\fontdimen2\font plus
\BIBentryALTinterwordstretchfactor\fontdimen3\font minus
  \fontdimen4\font\relax}
\providecommand{\BIBforeignlanguage}[2]{{%
\expandafter\ifx\csname l@#1\endcsname\relax
\typeout{** WARNING: IEEEtran.bst: No hyphenation pattern has been}%
\typeout{** loaded for the language `#1'. Using the pattern for}%
\typeout{** the default language instead.}%
\else
\language=\csname l@#1\endcsname
\fi
#2}}
\providecommand{\BIBdecl}{\relax}
\BIBdecl

\bibitem{pahlavanWidebandLocalAccess1997}
K.~Pahlavan, A.~Zahedi, and P.~Krishnamurthy, ``Wideband local access: Wireless
  {{LAN}} and wireless {{ATM}},'' \emph{IEEE Communications Magazine}, vol.~35,
  no.~11, pp. 34--40, Nov. 1997.

\bibitem{bianchiPerformanceEvaluationEnhancement1996}
G.~Bianchi, L.~Fratta, and M.~Oliveri, ``Performance evaluation and enhancement
  of the {{CSMA}}/{{CA MAC}} protocol for 802.11 wireless {{LANs}},'' in
  \emph{Proceedings of {{PIMRC}} '96 - 7th {{International Symposium}} on
  {{Personal}}, {{Indoor}}, and {{Mobile Communications}}}, vol.~2, Oct. 1996,
  pp. 392--396 vol.2.

\bibitem{pyattaevCommunicationChallengesHighdensity2015}
A.~Pyattaev, K.~Johnsson, S.~Andreev, and Y.~Koucheryavy, ``Communication
  challenges in high-density deployments of wearable wireless devices,''
  \emph{IEEE Wireless Communications}, vol.~22, no.~1, pp. 12--18, Feb. 2015.

\bibitem{ganjiCharacterizingPerformanceWiFi2019}
A.~Ganji, G.~Page, and M.~Shahzad, ``Characterizing the {{Performance}} of
  {{WiFi}} in {{Dense IoT Deployments}},'' in \emph{2019 28th {{International
  Conference}} on {{Computer Communication}} and {{Networks}} ({{ICCCN}})},
  Jul. 2019, pp. 1--9.

\bibitem{mengIEEE80211Real2018}
K.~Meng, A.~Jones, D.~Cavalcanti, K.~Iyer, C.~Ji, K.~Sakoda, A.~Kishida,
  F.~Hsu, J.~Yee, and L.~Li, ``{{IEEE}} 802.11 {{Real Time Applications TIG
  Report}},'' \emph{Doc: IEEE}, pp. 802--11, 2018.

\bibitem{wangStatisticalDelayErrorRate2024}
J.~Wang, W.~Cheng, and H.~Vincent~Poor, ``Statistical {{Delay}} and
  {{Error-Rate Bounded QoS Provisioning}} for {{AoI-Driven 6G Satellite-
  Terrestrial Integrated Networks Using FBC}},'' \emph{IEEE Transactions on
  Wireless Communications}, vol.~23, no.~10, pp. 15\,540--15\,554, Oct. 2024.

\bibitem{pan2024macrevivoartificialintelligence}
\BIBentryALTinterwordspacing
J.~Pan, J.~Wang, Z.~Yun, Z.~Xiao, Y.~Ouyang, W.~Cheng, and W.~Zhang, ``Mac
  revivo: Artificial intelligence paves the way,'' 2024. [Online]. Available:
  \url{https://arxiv.org/abs/2410.15820}
\BIBentrySTDinterwordspacing

\bibitem{zhangMultipleAccessIntegrated2024}
Y.~Zhang, W.~Cheng, and W.~Zhang, ``Multiple {{Access Integrated Adaptive
  Finite Blocklength}} for {{Ultra-Low Delay}} in {{6G Wireless Networks}},''
  \emph{IEEE Transactions on Wireless Communications}, vol.~23, no.~3, pp.
  1670--1683, Mar. 2024.

\bibitem{szottWifiMeetsML2022}
S.~Szott, K.~{Kosek-Szott}, P.~Gaw{\l}owicz, J.~T. G{\'o}mez, B.~Bellalta,
  A.~Zubow, and F.~Dressler, ``Wi-fi meets {{ML}}: {{A}} survey on improving
  {{IEEE}} 802.11 performance with machine learning,'' \emph{IEEE
  Communications Surveys \& Tutorials}, vol.~24, no.~3, pp. 1843--1893, 2022.

\bibitem{xiaTransmitPowerControl2015}
P.~Xia, Z.~Teng, and J.~Wu, ``Transmit power control and clear channel
  assessment in {{LAA}} networks,'' in \emph{2015 {{European Conference}} on
  {{Networks}} and {{Communications}} ({{EuCNC}})}, Jun. 2015, pp. 210--213.

\bibitem{ramachandranClearChannelAssessment2007}
I.~Ramachandran and S.~Roy, ``Clear channel assessment in energyconstrained
  wideband wireless networks,'' \emph{IEEE Wireless Communications}, vol.~14,
  no.~3, pp. 70--78, Jun. 2007.

\bibitem{dengIEEE80211beWiFi2020}
C.~Deng, X.~Fang, X.~Han, X.~Wang, L.~Yan, R.~He, Y.~Long, and Y.~Guo,
  ``{{IEEE}} 802.11be {{Wi-Fi}} 7: {{New Challenges}} and {{Opportunities}},''
  \emph{IEEE Communications Surveys \& Tutorials}, vol.~22, no.~4, pp.
  2136--2166, 2020.

\bibitem{kosek-szottCoexistenceIssuesFuture2017}
K.~{Kosek-Szott}, J.~Gozdecki, K.~Loziak, M.~Natkaniec, L.~Prasnal, S.~Szott,
  and M.~Wagrowski, ``Coexistence {{Issues}} in {{Future WiFi Networks}},''
  \emph{IEEE Network}, vol.~31, no.~4, pp. 86--95, Jul. 2017.

\bibitem{DynamicSensitivityControl}
``Dynamic sensitivity control of access points for {{IEEE}} 802.11ax {\textbar}
  {{IEEE Conference Publication}} {\textbar} {{IEEE Xplore}},''
  https://ieeexplore.ieee.org/abstract/document/7511025.

\bibitem{lvAdaptiveRateCarrier2017}
Z.~Lv, H.~Hu, D.~Yuan, and J.~Ran, ``An adaptive rate and carrier sense
  threshold algorithm to enhance throughput and fairness for dense {{WLANs}},''
  in \emph{2017 3rd {{IEEE International Conference}} on {{Computer}} and
  {{Communications}} ({{ICCC}})}, Dec. 2017, pp. 453--458.

\bibitem{huaDistributedPhysicalCarrier2009}
Y.~Hua, Q.~Zhang, and Z.~Niu, ``Distributed {{Physical Carrier Sensing
  Adaptation Scheme}} in {{Cooperative MAP WLAN}},'' in \emph{{{GLOBECOM}} 2009
  - 2009 {{IEEE Global Telecommunications Conference}}}, Nov. 2009, pp. 1--6.

\bibitem{zhuOptimalPhysicalCarrier2007}
Y.~Zhu, Q.~Zhang, Z.~Niu, and J.~Zhu, ``On {{Optimal Physical Carrier
  Sensing}}: {{Theoretical Analysis}} and {{Protocol Design}},'' in
  \emph{{{IEEE INFOCOM}} 2007 - 26th {{IEEE International Conference}} on
  {{Computer Communications}}}, May 2007, pp. 2351--2355.

\bibitem{afaquiEvaluationDynamicSensitivity2015}
M.~S. Afaqui, E.~{Garcia-Villegas}, E.~{Lopez-Aguilera}, G.~Smith, and
  D.~Camps, ``Evaluation of dynamic sensitivity control algorithm for {{IEEE}}
  802.11ax,'' in \emph{2015 {{IEEE Wireless Communications}} and {{Networking
  Conference}} ({{WCNC}})}, Mar. 2015, pp. 1060--1065.

\bibitem{kimFACTFineGrainedAdaptation2017}
S.~Kim, S.~Yoo, J.~Yi, Y.~Son, and S.~Choi, ``{{FACT}}: {{Fine-Grained
  Adaptation}} of {{Carrier Sense Threshold}} in {{IEEE}} 802.11 {{WLANs}},''
  \emph{IEEE Transactions on Vehicular Technology}, vol.~66, no.~2, pp.
  1886--1891, Feb. 2017.

\bibitem{kimImprovingSpatialReuse2006}
T.-S. Kim, H.~Lim, and J.~C. Hou, ``Improving spatial reuse through tuning
  transmit power, carrier sense threshold, and data rate in multihop wireless
  networks,'' in \emph{Proceedings of the 12th Annual International Conference
  on {{Mobile}} Computing and Networking}, ser. {{MobiCom}} '06.\hskip 1em plus
  0.5em minus 0.4em\relax New York, NY, USA: Association for Computing
  Machinery, Sep. 2006, pp. 366--377.

\bibitem{fuemmelerSelectingTransmitPowers2006}
J.~A. Fuemmeler, N.~H. Vaidya, and V.~V. Veeravalli, ``Selecting transmit
  powers and carrier sense thresholds in {{CSMA}} protocols for wireless ad hoc
  networks,'' in \emph{Proceedings of the 2nd Annual International Workshop on
  {{Wireless}} Internet}, ser. {{WICON}} '06.\hskip 1em plus 0.5em minus
  0.4em\relax New York, NY, USA: Association for Computing Machinery, Aug.
  2006, pp. 15--es.

\bibitem{nakashimaDeepReinforcementLearningBased2020}
K.~Nakashima, S.~Kamiya, K.~Ohtsu, K.~Yamamoto, T.~Nishio, and M.~Morikura,
  ``Deep {{Reinforcement Learning-Based Channel Allocation}} for {{Wireless
  LANs With Graph Convolutional Networks}},'' \emph{IEEE Access}, vol.~8, pp.
  31\,823--31\,834, 2020.

\bibitem{choReinforcementLearningRate2021}
S.~Cho, ``Reinforcement {{Learning}} for {{Rate Adaptation}} in {{CSMA}}/{{CA
  Wireless Networks}},'' in \emph{Advances in {{Computer Science}} and
  {{Ubiquitous Computing}}}, J.~J. Park, S.~J. Fong, Y.~Pan, and Y.~Sung,
  Eds.\hskip 1em plus 0.5em minus 0.4em\relax Singapore: Springer, 2021, pp.
  175--181.

\bibitem{cakirDTWNQlearningbasedTransmit2022}
L.~V. {\c C}ak{\i}r, K.~Huseynov, E.~Ak, and B.~Canberk, ``{{DTWN}}:
  {{Q-learning-based Transmit Power Control}} for {{Digital Twin WiFi
  Networks}},'' \emph{EAI Endorsed Transactions on Industrial Networks and
  Intelligent Systems}, vol.~9, no.~31, pp. e5--e5, Jun. 2022.

\bibitem{huangDeepQNetworkApproach2022}
Y.~Huang and K.-W. Chin, ``A {{Deep Q-Network Approach}} to {{Optimize Spatial
  Reuse}} in {{WiFi Networks}},'' \emph{IEEE Transactions on Vehicular
  Technology}, vol.~71, no.~6, pp. 6636--6646, Jun. 2022.

\bibitem{huangThreeTierDeepLearningBased2023}
------, ``A {{Three-Tier Deep Learning-Based Channel Access Method}} for {{WiFi
  Networks}},'' \emph{IEEE Transactions on Machine Learning in Communications
  and Networking}, vol.~1, pp. 90--106, 2023.

\bibitem{goldsmithWirelessCommunications}
A.~Goldsmith, \emph{Wireless {{Communications}}}.\hskip 1em plus 0.5em minus
  0.4em\relax Cambridge, UK: Cambridge University Press, 2005.

\bibitem{cagaljSelfishBehaviorCSMA2005}
M.~Cagalj, S.~Ganeriwal, I.~Aad, and J.-P. Hubaux, ``On selfish behavior in
  {{CSMA}}/{{CA}} networks,'' in \emph{Proceedings {{IEEE}} 24th {{Annual Joint
  Conference}} of the {{IEEE Computer}} and {{Communications Societies}}.},
  vol.~4, Mar. 2005, pp. 2513--2524 vol. 4.

\bibitem{IEEEStandardInformation2011}
``{{IEEE}} standard for information technology--telecommunications and
  information exchange between systems--local and metropolitan area
  networks--specific requirements part 11: {{Wireless LAN}} medium access
  control ({{MAC}}) and physical layer ({{PHY}}) specifications amendment 10:
  Mesh networking,'' \emph{IEEE Std 802.11s-2011 (Amendment to IEEE Std
  802.11-2007 as amended by IEEE 802.11k-2008, IEEE 802.11r-2008, IEEE
  802.11y-2008, IEEE 802.11w-2009, IEEE 802.11n-2009, IEEE 802.11p-2010, IEEE
  802.11z-2010, IEEE 802.11v-2011, and IEEE 802.11u-2011)}, pp. 1--372, Sep.
  2011.

\bibitem{zhao2024RLBook}
S.~Zhao, \emph{Mathematical Foundations of Reinforcement Learning}.\hskip 1em
  plus 0.5em minus 0.4em\relax Springer Nature Press, 2024.

\bibitem{yinNs3aiFosteringArtificial2020}
H.~Yin, P.~Liu, K.~Liu, L.~Cao, L.~Zhang, Y.~Gao, and X.~Hei, ``Ns3-ai:
  {{Fostering Artificial Intelligence Algorithms}} for {{Networking
  Research}},'' in \emph{Proceedings of the 2020 {{Workshop}} on Ns-3}, ser.
  {{WNS3}} '20.\hskip 1em plus 0.5em minus 0.4em\relax New York, NY, USA:
  Association for Computing Machinery, Jun. 2020, pp. 57--64.

\bibitem{lecciNs3ImplementationBursty2021}
M.~Lecci, A.~Zanella, and M.~Zorzi, ``An ns-3 implementation of a bursty
  traffic framework for virtual reality sources,'' in \emph{Proceedings of the
  {{Workshop}} on Ns-3}, ser. {{WNS3}} '21.\hskip 1em plus 0.5em minus
  0.4em\relax Virtual Event USA: ACM, Jun. 2021, pp. 73--80.

\end{thebibliography}

\vfill

\end{document}